\documentclass[epj,final]{svjour}[07-02-2011]
\usepackage{epsfig}
\usepackage{amssymb}
\usepackage{amsmath}
\usepackage{graphicx}
\usepackage{epstopdf}
\newcommand{\bm}[1]{ \mbox{\boldmath $#1$}  }

\begin{document}

\title{Density Waves in Layered Systems with Fermionic Polar Molecules}

\author{N.~T. Zinner \and G.~M. Bruun}
  \institute{Department of Physics and Astronomy - Aarhus University, Ny Munkegade, bygn. 1520, DK-8000 \AA rhus C, Denmark}
\date{\today}

\abstract{
A layered system of two-dimensional planes containing fermionic polar molecules can potentially realize a 
number of exotic quantum many-body states. Among the predictions, are density-wave instabilities driven by 
the anisotropic part of the dipole-dipole interaction in a single layer. However, in typical 
multilayer setups it is reasonable to expect that the onset and properties of a density-wave are 
modified by adjacent layers. Here we show that this is indeed the case. For multiple layers the 
critical strength for the density-wave instability decreases with the number of layers. The effect depends on density and 
is more pronounced in the low density regime. The lowest solution of the instability corresponds to the 
density waves in the different layers being in-phase, whereas higher solutions have one
or several adjacent layers that are out of phase.
The parameter regime needed to explore this instability is within reach of current experiments.
\PACS{
{03.75.Ss}{Degenerate Fermi Gases} 
\and  
{05.30.Fk}{Fermion systems and electron gas} \and
{67.85.-d}{Ultracold gases, trapped gases} }
}
\maketitle

\section{Introduction}
After the  great successes of cold atomic gas physics using neutral atoms with short-range interaction \cite{bloch2008,giorgini2008}, 
many groups
have now set their goals on obtaining ultracold samples of polar molecules that have an anisotropic long-range interaction~\cite{doyle2004,ospelkaus2008,ni2008,deiglmayr2008,lang2008,ospelkaus2010,ni2010,miranda2010}. These can, however, lead to strong losses and the design of experimental geometries that reduce these effects are
now becoming a reality. In particular, the use of two-dimensional geometries can reduce losses and at the same time very
interesting many-body phases in both single- and multilayer configurations have been proposed \cite{baranov2005,wang2006,wang2007,mora2007,buchler2007,astra2007,wang2008,bruun2008,baranov2008,shih2009,lahaye2009,dsarma2009,cooper2009,potter2010,pikovski2010,zinner2010,baranov2011}.
One such proposal concerns the potential instability of a single two-dimensional layer with polar fermions toward the 
formation of density-waves as the polarization of the molecules with respect to the layer plane is varied \cite{yamaguchi2010,sun2010}.
However, the systems of current experimental interest are not single-layer \cite{miranda2010}, and the effect of
adjacent layers is therefore of concern. 

Using linear response within the  the random-phase approximation,
 we consider how interlayer interactions influence
the density-waves instability and how the critical strength is modified by
interlayer terms. In order to estimate the effects of exchange terms, we use many-body local field factors. This approach has been  successfully applied to electron systems. 
We find that the instability is enhanced by the presence of in-phase density-waves in neighboring layers. 
The effect depends on the 
density of fermions in each layer and is most pronounced in the low density limit where the critical value is inversely
proportional to the number of layers. The latter effect is largely insensitive to the inclusion of exchange terms, Fermi surface deformation, 
or changes in the effective mass. The density-wave instability will therefore occupy a larger region of the
zero-temperature phase diagram for a multilayered system as compared to a single layer system. 

\section{Linear Response and Effective Interaction}
\begin{figure}[t]
\begin{center}
\leavevmode
\begin{minipage}{1\columnwidth}
\includegraphics[clip=true,height=0.6\columnwidth,width=1\columnwidth]{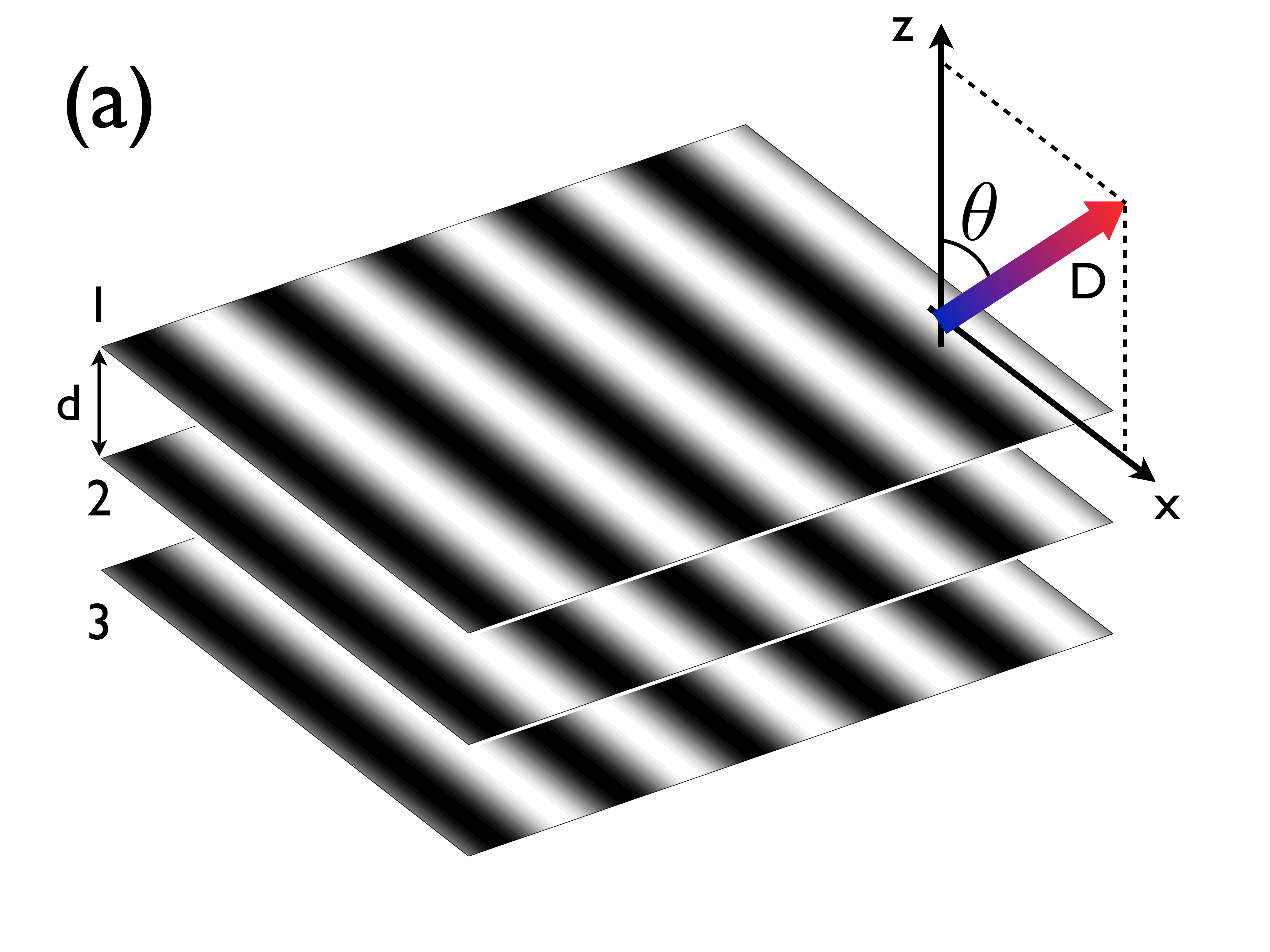}
\end{minipage}
\begin{minipage}{1\columnwidth}
\includegraphics[clip=true,height=0.6\columnwidth,width=1\columnwidth]{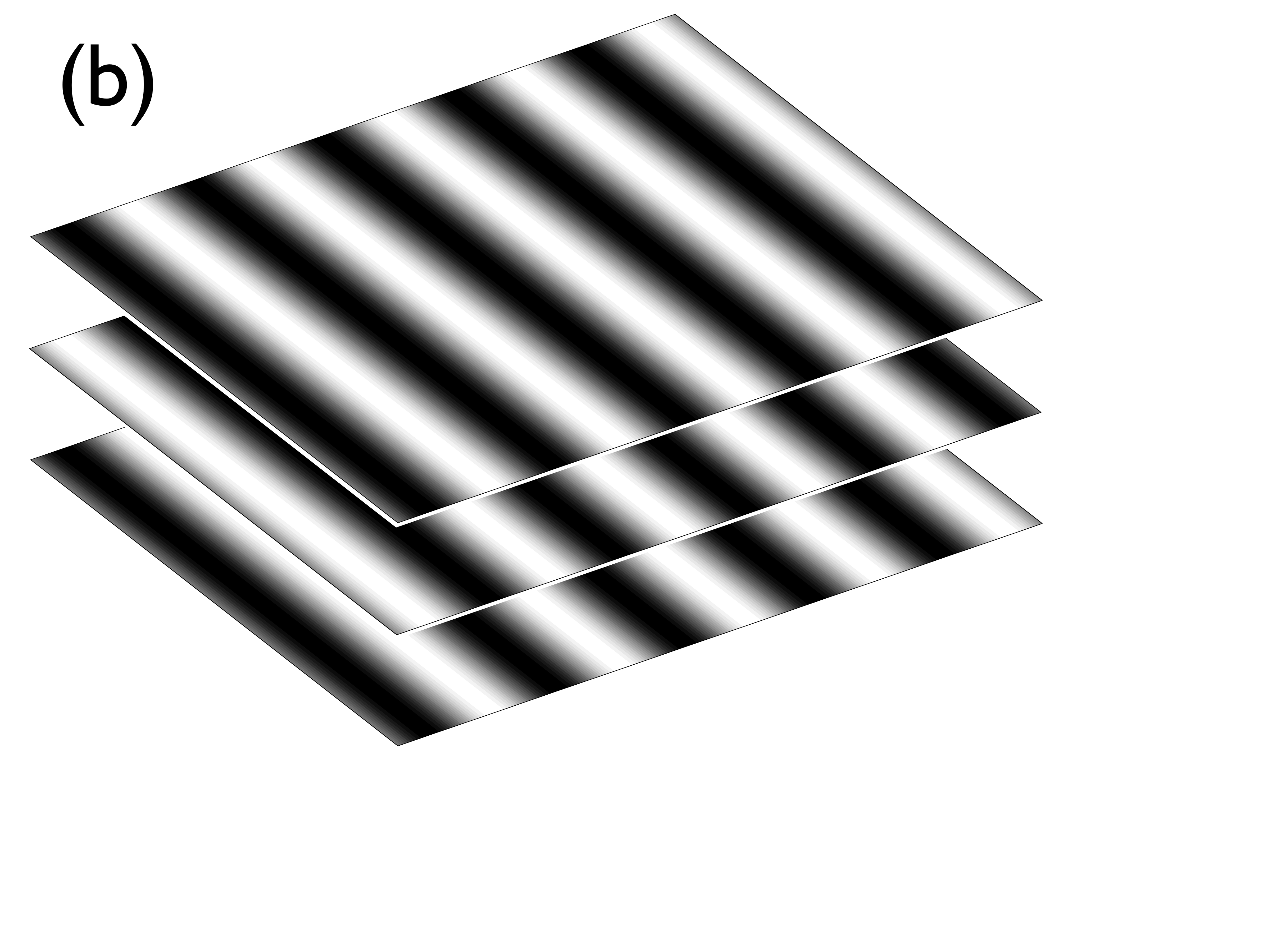}
\end{minipage}
\begin{minipage}{1\columnwidth}
\includegraphics[clip=true,height=0.6\columnwidth,width=1\columnwidth]{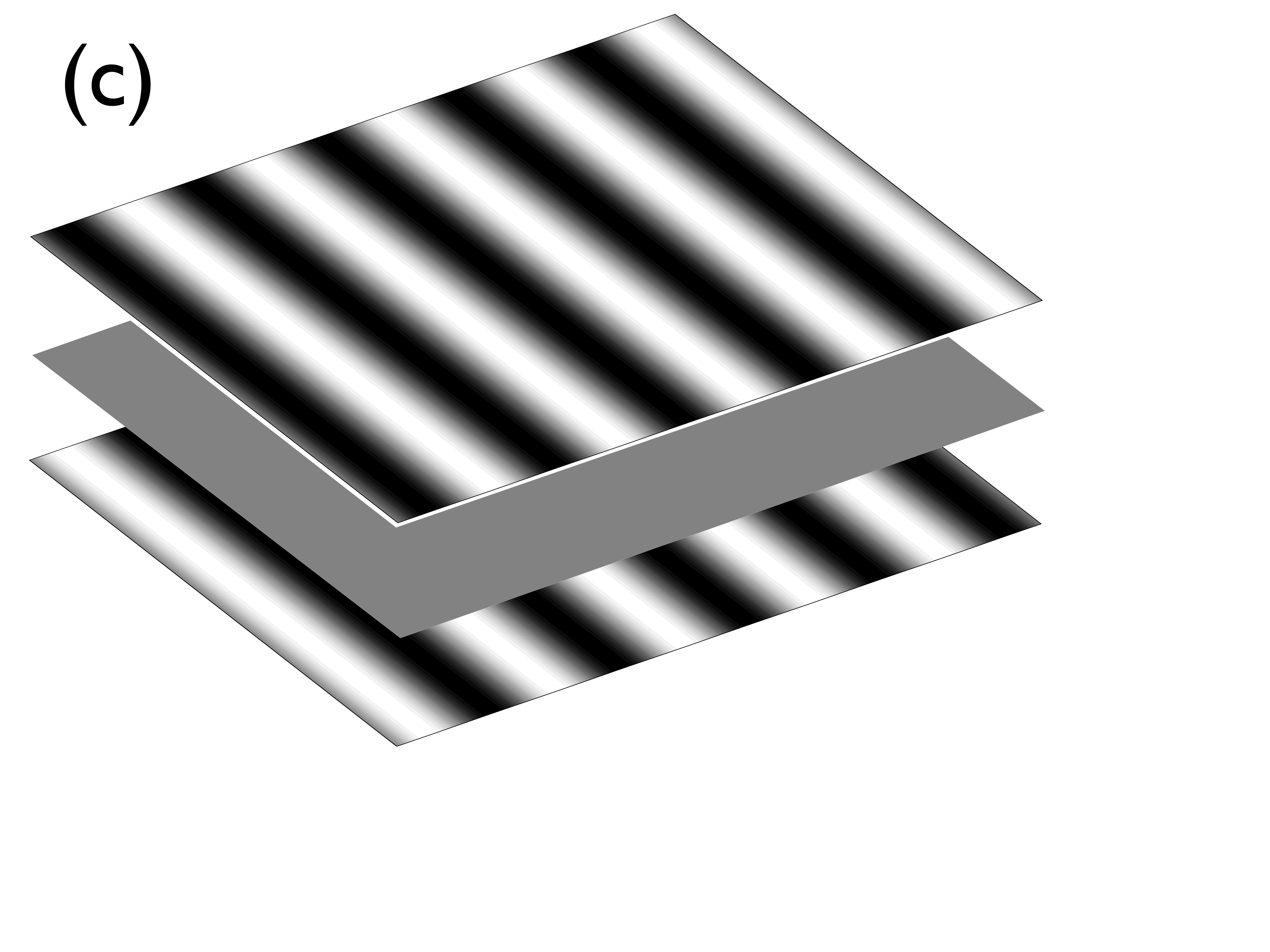}
\end{minipage}
\caption{The experimental setup and the density waves corresponding to the lowest (a), the next lowest (b), and the highest (c) eigenmodes for the three layer case. 
The dipole moments form the angle $\theta$ with respect to the normal of the planes. Their projection onto the planes is parallel to the wave fronts.}
  \label{Cartoon}
\end{center}
\end{figure}

We consider a multilayer system of fermions with dipole moment ${\mathbf D}$ and mass $m$ confined in planes parallel to the $xy$-plane
and separated by the distance $d$. In the direction normal to the planes, all dipoles reside in the lowest 
quantum level which we take to be a Gaussian of width $w$, i.e.\
 $\phi(z)\propto \exp(-z^2/2w^2)$. The dipole moments ${\mathbf D}$ are aligned by an external field forming an angle $\theta$ with respect 
 to the normal of the planes and with a projection onto the planes which is parallel to the $x$-axis. The experimental setup is 
 illustrated in Fig.~\ref{Cartoon}. 
Two dipoles separated by ${\mathbf r}$ interact with the potential $V({\mathbf r})=D^2(1-3\cos^2\theta_{\rm rd})/r^3$
where 
$\theta_{\rm rd}$ is the angle between ${\mathbf D}$ and ${\mathbf r}$.
We assume that the layers all have the same  density $n$ of fermions.

To obtain the instabilities of the multilayered system we use linear response theory and the random-phase 
approximation (RPA) as was done for the case of a single layer in \cite{yamaguchi2010,sun2010}. 
Within the RPA framework, the density-wave 
instability occurs at the poles of the density-density response function. To treat several layers we   
extend the RPA to a multilayer (or multicomponent) system. We can write a general density fluctuation in 
response to an external potential,  $\phi_{ex}$, in momentum ($\bm q$) and frequency ($\omega$) space as
\begin{align}\label{fullrho}
\delta {\bm\rho}(\bm q,\omega)=\bm \chi(\bm q,\omega){\bm\phi}_{ex}(\bm q,\omega),
\end{align}
where $\delta{\bm\rho}$ is a vector quantity containing the disturbances in each layer as entries. Likewise,
$\bm \chi(\bm q,\omega)$ is now in general a matrix of response function with entries $\chi_{ij}(\bm q,\omega)$. 
The interactions between the various layers produce an induced potential which we write as
\begin{align}
{\bm\phi}_{ind}(\bm q,\omega)=\bm V(\bm q) \delta {\bm\rho}(\bm q,\omega),
\end{align}
where the matrix $\bm V(\bm q)_{ij}=V_{ij}(\bm q)$ contains the interaction between layers $i$ and $j$. We
approximate the system response to that of a non-interacting Fermi gas responding to both the external and 
the induced disturbance. We then have 
\begin{align}\label{apprho}
\delta {\bm\rho}(\bm q,\omega)=
\bm \chi^0(\bm q,\omega)\left[{\bm\phi}_{ex}(\bm q,\omega)+\bm V(\bm q) \delta {\bm\rho}(\bm q,\omega)\right],
\end{align}
where $\bm \chi^0(\bm q,\omega)_{ij}=\delta_{ij}\chi^{0}_{i}(\bm q,\omega)$ is the matrix of response functions
of the non-interacting system which is of course diagonal. Combining Eqs.~\eqref{fullrho} and \eqref{apprho}
we arrive at the following matrix equation the response function
\begin{align}\label{RPA}
\bm \chi(\bm q,\omega)=\left[I-\bm \chi^{0}(\bm q,\omega)\bm V(\bm q)\right]^{-1}\bm \chi^{0}(\bm q,\omega).
\end{align}
In the case of a single layer this equation reduces to the standard RPA expression for the density-density 
response function. Here we are interesed in density-wave instabilities in the static limit $\omega=0$ and 
we have to determine the singularities of $\bm \chi(\bm q)$. By inversion, we see that these occur when 
\begin{align}\label{det}
\text{det}[I-\bm \chi^{0}(\bm q)\bm V(\bm q)]=0,
\end{align}
and this is the equation that we will solve below.

We assume here that the density in each layer is the same,  so that the non-interacting response functions are
all the same, i.e. $\chi^{0}_{i}(\bm q)=\chi^0(\bm q)$, and are given by
\begin{align}\label{response}
\chi^0(\bm q)=\int\frac{d^2\bm k}{(2\pi)^2}\frac{f(\bm k+\bm q)-f(\bm k)}{\epsilon_{\bm k+\bm q}-\epsilon_{\bm k}},
\end{align}
where $\epsilon_{\bm k}=\hbar^2\bm k^2/2m$ and $f$ is the Fermi distribution. In the two-dimensional case of interest here
we have the explicit expression \cite{stern1967}
\begin{align}
\chi^0(\bm q)=\frac{m}{2\pi\hbar^2}\left[\sqrt{1-\left(\frac{2k_F}{q}\right)^2}\theta(q-2k_F)-1\right],
\end{align}
where $k_F$ is the Fermi momentum and $\theta(x)$ is the Heaviside step-function. For simplicity, we ignore any Fermi surface 
deformation  due to the dipolar interaction~\cite{yamaguchi2010}.
We will briefly comment on the influence of such effects  in Sec.~\ref{Deformation}.

\subsection{Exchange Corrections}
The RPA analysis above neglects the role of exchange interactions. In the single-component Fermi system 
we consider here, the exchange effect can be significant. As an example, we note that for a momentum-independent
potential, the exchange correction would completely cancel the direct term in a Hartree-Fock calculation. 
The effective dipolar interaction that we discuss in the next section depends, however, linearly on momentum. 
The effects of exchange can be included via the Hartree-Fock RPA approximation. This unfortunately involves a non-local interaction making the resulting 
numerics somewhat involved. 
We will not pursue such calculations here, but rather follow the simpler
local field factor approach that has been very successful for the electron liquid \cite{giuli2005}. 
It attempts to include the intrinsically non-local effects of the exchange term
through the introduction of an effective local 'exchange' potential, in similar
spirit to the highly successful density-functional method.
This approach has been applied to two-dimensional double-layer electron systems (see for example Ref. \cite{zheng1994})
which is a system closely analogous to the one studied here.

In the multilayer setup considered here, we must be careful when including exchange corrections in the 
correct places. Since we assume that there is no tunneling between the layers, the layer index is 
effectively a spin coordinate, and we therefore have {\it no} exchange corrections for the interlayer
interaction. This means that only the diagonal terms in Eq.~\eqref{RPA} have to be modified. We write a diagonal entry in the form
\begin{equation}
1-V_0(\bm q)\left[1-G(\bm q)\right]\chi^0(\bm q),
\end{equation} 
where $G(\bm q)$ is the momentum-dependent local field factor. 
There are various more or less sophisticated
ways to calculate this factor through self-consistent numerical methods \cite{giuli2005}. As 
we are only interested in estimating the effects of exchange correlations on the density wave 
instability, we will follow a more intuitive approach originally introduced by Hubbard \cite{hubbard1957}. 

The Pauli principle introduces the so-called 'exchange-hole' in Fermi systems. For large $\bm q$, i.e. short length scales,  the exchange-hole  cancels 
the direct interaction and $G(\bm q)\rightarrow 1$
for $q\rightarrow \infty$. For $q\rightarrow 0$, i.e. for long distance, the exchange effect
should not play a role and in turn $G(\bm q)\rightarrow 0$. Between these limits, the detailed functional form of 
$G(\bm q)$ of course depends on the particular form of the bare potential $V_0(\bm q)$. 
As we are only interested in the qualitative effects
of the exchange correlations, it is  sufficient to use the  simple function $G(\bm q)=\tfrac{2}{\pi}\,\textrm{tan}^{-1}(q/s)$ 
which interpolates between the $q=0$ and $q\rightarrow \infty$  limits above.  Here $s$ is
the natural scale in the problem at hand; we take $s=2k_F$.  

\subsection{Effective Dipolar Interaction}
The direct dipole-dipole interaction has an intra- and an interlayer part in our multilayered setup. The Fourier transform of the 
former can be written \cite{fischer2006}
\begin{align}\label{intra}
V_0(\bm q)=\frac{4\pi D^2}{\sqrt{2\pi}w}\left[\frac{2}{3}P_2(\cos\theta)-\xi(\theta,\alpha)F(qw)\right],
\end{align}
where $q=|\bm q|$ and $\alpha$ is the azimuthal angle between the wave vector ${\mathbf q}=(q_x,q_y)$ and the projection of ${\mathbf D}$ onto the plane
which is parallel to the $x$-axis.  $P_2(x)$
is the second Legendre polynomial, and we have defined the function $F(x)=\sqrt{\tfrac{\pi}{2}}x[1-\textrm{erf}(x/\sqrt{2})]\exp(x^2/2)$ with
$\textrm{erf}(x)$ the error function. To obtain this formula, the $z$-direction confining the dipoles in the layers have been integrated
out. The interesting angular dependence
of the intralayer interaction is contained in the function $\xi(\theta,\alpha)=\cos^2\theta-\sin^2\theta\cos^2\alpha$.
This function provides the anisotropy in momentum space which is absent at $\theta=0$ 
when the dipoles are oriented perpendicularly to the layer.
For $w\ll d$, the
interlayer interaction can be written as~\cite{wang2008}
\begin{align}
V_1(\bm q)=-2\pi D^2\xi(\theta,\alpha) q e^{-d q}.
\end{align}
This  approximation  holds very well for small $w$ and deviates less than 10\% for $w=0.2d$.

As argued in \cite{yamaguchi2010,sun2010}, the most unstable direction is found at  
$\alpha=\pi/2$. This is a configuration where the density-wave is perpendicular to 
the $x$-axis in order to reduce the side-by-side repulsion of the dipoles while optimizing the attraction from the head-to-tail setup, see Fig.~\ref{Cartoon}. In this case we have $\xi=\cos^2\theta$. 
The first term in Eq.~\eqref{intra} which is constant in momentum space can be discarded 
since we are working with a single-component Fermi system \cite{yamaguchi2010,sun2010}.
As discussed in \cite{yamaguchi2010}, the critical value in a single-layer has some dependence on $\theta$. Here we are interesting in the effects of multiple layers and we thus fix $\theta$ at $\cos^2\theta=1/3$, but 
our results can be easily mapped to a different angle through the substitution $D^2/3\rightarrow D^2 \cos^2\theta$.

With the choices above, the intralayer interaction becomes
\begin{align}\label{vpot}
V_0(q)=-\frac{4\pi D^2}{3\sqrt{2\pi}w}F(qw),
\end{align}
whereas the interlayer interaction is simply multiplied by a factor of $\tfrac{1}{3}$. For $w q\ll 1$, Eq.~\eqref{vpot} reproduces the
potential used for the single layer case in \cite{yamaguchi2010,sun2010}. As we are mostly concerned with the effects of multiple layers, 
we will also assume $wq\ll 1$ for simplicity, i.e. we assume that $V_0(q)$ is linear in $q$. This linear momentum dependence of the intralayer potential was used in \cite{sun2010} to argue that the density-wave instability must occur at some dipole strength always. We note that the most
unstable mode is expected to be at $q=2k_F$
(neglecting the effects of Fermi surface deformation). For consistency, we must therefore have that $2k_Fw\ll 1$. In terms of the 
density of a single layer, $n$, this condition reads $w\sqrt{16\pi n}\ll 1$. Thus, either the density must be small or the transverse
confinement strong. In terms of typical physical scales in experiments, we have 
\begin{align}
2k_Fw=7.1\frac{w}{1\mu \textrm{m}}\sqrt{\frac{n}{10^8 \textrm{cm}^{-2}}}.
\end{align}
Using experimentally relevant values \cite{miranda2010} $d=0.5\mu\textrm{m}$ and assuming $w/d=0.1$, we find that $2k_Fw<1$ for densities $n<8\cdot 10^8$ cm$^{-2}$ which is fulfilled by current experiments.

\section{Instability Conditions}
We first recapitulate the findings for a single layer in the 
RPA neglecting exchange. The instability equation 
is 
\begin{align}
1-\chi^0(q)V_0(q)=0.
\end{align}
Assuming that the instability occurs first at $q=2k_F$, we have $\chi^0=-\tfrac{m}{2\pi\hbar^2}$ and $V_0(2k_F)=-4\pi D^2k_F/3$.
We thus have the relation 
\begin{align}
D_{c}^{2}=\frac{3\hbar^2}{2mk_F},
\end{align}
where $D_c$ is the critical dipole strength \cite{sun2010}. We define a dimensionless measure of the
strength and density; $g:=2mD^2k_F/3\hbar^2$. We thus have the critical value $g_0=1$. To include exchange, we need to make 
the substitution $V_0(2k_F)\rightarrow V_0(2k_F)[1-G(2k_F)]=V_0(2k_F)/2$ and we obtain $g_0=2$ instead. 
To highlight the effects of the multilayer setup, we now proceed to 
discuss the bi- and trilayer cases without the $1-G(q)$ factors and defer the discussion of exchange corrections to Sec.~\ref{exchange}.

\subsection{The Bilayer}
For the case of two adjacent layers we get the following algebraic equation from Eq.~\eqref{det}
\begin{align}
\left[1-\chi^0(q)V_0(q) \right]^2-\left[\chi^0(q)V_1(q)\right]^2=0.
\end{align}
We note immediately that if we set $V_1(q)=0$ we recover the usual RPA condition for density instabilities. It is 
also clear at this point that the bilayer will have a smaller $D_c$ than the single layer above since the 
$V_1(q)$ term is negative. If one considers the interlayer interaction from the point of view of
induced interactions this is no surprise as such interaction are usually attractive at lowest order.

At $q=2k_F$ we can solve the equation above and find the lowest critical value for a density-wave instability in a bilayer
\begin{align}
g_b=\frac{1}{1+e^{-2k_Fd}}<1=g_0.
\end{align}
 For $k_Fd=1$, we get roughly a 12 percent reduction, whereas for 
a lower density of $k_Fd=0.5$ the difference is 27 percent. The other solution to the bilayer equation
is $\tilde g_b=(1-e^{-2k_Fd})^{-1}$, so that $g_b<1<\tilde g_b$ for all $k_Fd$. Solving for the corresponding 
zero eigenmodes of $\bm \chi(\bm q,0)^{-1}$ from Eq.~\eqref{RPA}, we find
\begin{align}
\begin{bmatrix}
\delta\rho_1\\ \delta\rho_2
\end{bmatrix}=
\left[\begin{matrix}1\\1\end{matrix}\right]\,\,\text{for}\,\,g_b\,\,\text{and}\,\,
\begin{bmatrix}
\delta\rho_1\\ \delta\rho_2
\end{bmatrix}=\left[\begin{matrix}1\\-1\end{matrix}\right]\,\,\text{for}\,\,\tilde g_b.
\end{align}
Here, $\delta\rho_i=\delta\rho_i(2k_F)$ is the density fluctuation in layer $i$. We see that  
the density waves in the two layers are in-phase for the lower solution $g_b$ and
out of phase for the $\tilde g_b$ solution. Thus, the instability is enhanced by the density waves in neighboring layers 
being in-phase  gaining more attractive head-to-tail energy and minimizing  the side-by-side repulsion. Likewise, 
when the density-waves are out-of-phase the instability is suppressed. 

\subsection{The Trilayer}
The case of three layers produces the algebraic equation
\begin{align}
&0=\left(1-\chi^0(q)V_0(q)\right)\left[(1-\chi^0(q)V_0(q))^2\right.&\nonumber\\
&\left.-2(\chi^0(q)V_1(q))^2-(\chi^0(q)V_2(q))^2\right]&\nonumber\\
&+2(\chi^0(q)V_1(q))^2\chi^0(q)V_2(q),&
\end{align}
where we have introduced the notation $V_2(q)$ for the interlayer potential of the two outer layers that
are a distance $2d$ apart. 
Note that  $V_2(q)$ differs from $V_1(q)$ by a factor of $\exp(-qd)$ and 
we thus expect it to be a much smaller quantity than $V_1(q)$ at $2k_F$.

In light of the above, we therefore first consider the simpler case of $V_2(q)=0$, i.e. we include only
nearest-neighbor interactions. This means that the equation for the instability factorizes and the
condition becomes that either $\left(1-\chi^0(q)V_0(q)\right)=0$ or
\begin{align}
\left[(1-\chi^0(q)V_0(q))^2-2(\chi^0(q)V_1(q))\right]=0.
\end{align}
Clearly the latter condition produces a lower critical value and we find for the trilayer with
$V_2(q)=0$ that
\begin{align}
g_{t}^{*}=\frac{1}{1+\sqrt{2}e^{-2k_Fd}}<g_b,
\end{align}
where the asterisk indicates that we include nearest-neighbor interactions only.
As expected the trilayer has a reduced critical value. A naive guess 
for the trilayer might be to multiply the interlayer strength by a factor of 2 and then consider 
it as a bilayer problem. However, our result demonstrates that the enhancement is only by
a factor of $\sqrt{2}$. As it turns out the trilayer equation has a rather simple analytic solution. 
The roots are 
\begin{align}
g_t=
\begin{cases}
\frac{1}{2}\frac{a^2+2-\sqrt{a^4+8a^2}}{1-a^2}\\
\frac{1}{2}\frac{a^2+2+\sqrt{a^4+8a^2}}{1-a^2}\\
\frac{1}{1-a^2}
\end{cases},
\end{align}
where $a=\exp(-2k_Fd)$ and we have listed 
 them in order of increasing  magnitude.
 The top solution is always less than one and decreases with $k_Fd$ (we denote it by $g_t$ in the following) whereas the others are always larger than one and increase with $k_Fd$. The corresponding eigenmodes are
 \begin{align}
\begin{bmatrix}
\delta\rho_1\\ \delta\rho_2\\ \delta\rho_3
\end{bmatrix}=
\left[\begin{matrix}1 \\ \frac{\sqrt{a^2+8}-a}{2}\\1 \end{matrix}\right]\, ,
\left[\begin{matrix}1 \\ -\frac{\sqrt{a^2+8}+a}{2}\\1 \end{matrix}\right]\, ,
\left[\begin{matrix}1 \\ 0\\-1 \end{matrix}\right]
\end{align}
where layer $2$ is the one in the middle, see Fig.~\ref{Cartoon}.  Again, the 
lowest solution corresponds to the density-waves in the different layers being in-phase with  amplitude now being the largest 
for the layer in the middle.  The second eigenmode has the middle layer out of phase
and of larger magnitude than the outer layers. This is the same situation as the $\tilde g_b$ solution for the bilayer, only
now the out of phase effect is more costly. This is also reflected in the fact that this solution is always larger than $\tilde g_b$
for any $k_Fd$, whereas the opposite holds for the lowest solution, i.e. $g_t<g_b$. The last solution is an eigenmode
with density waves of the outer layers out of phase and no amplitude change in the middle layer (within the RPA). 
We sketch in Fig.\ \ref{Cartoon} the density waves for the three eigenmodes.
The physically relevant mode for the instability  is of course the one with the lowest critical value corresponding to 
the density waves in the planes being in-phase.  

\begin{figure}[h]
  \epsfig{file=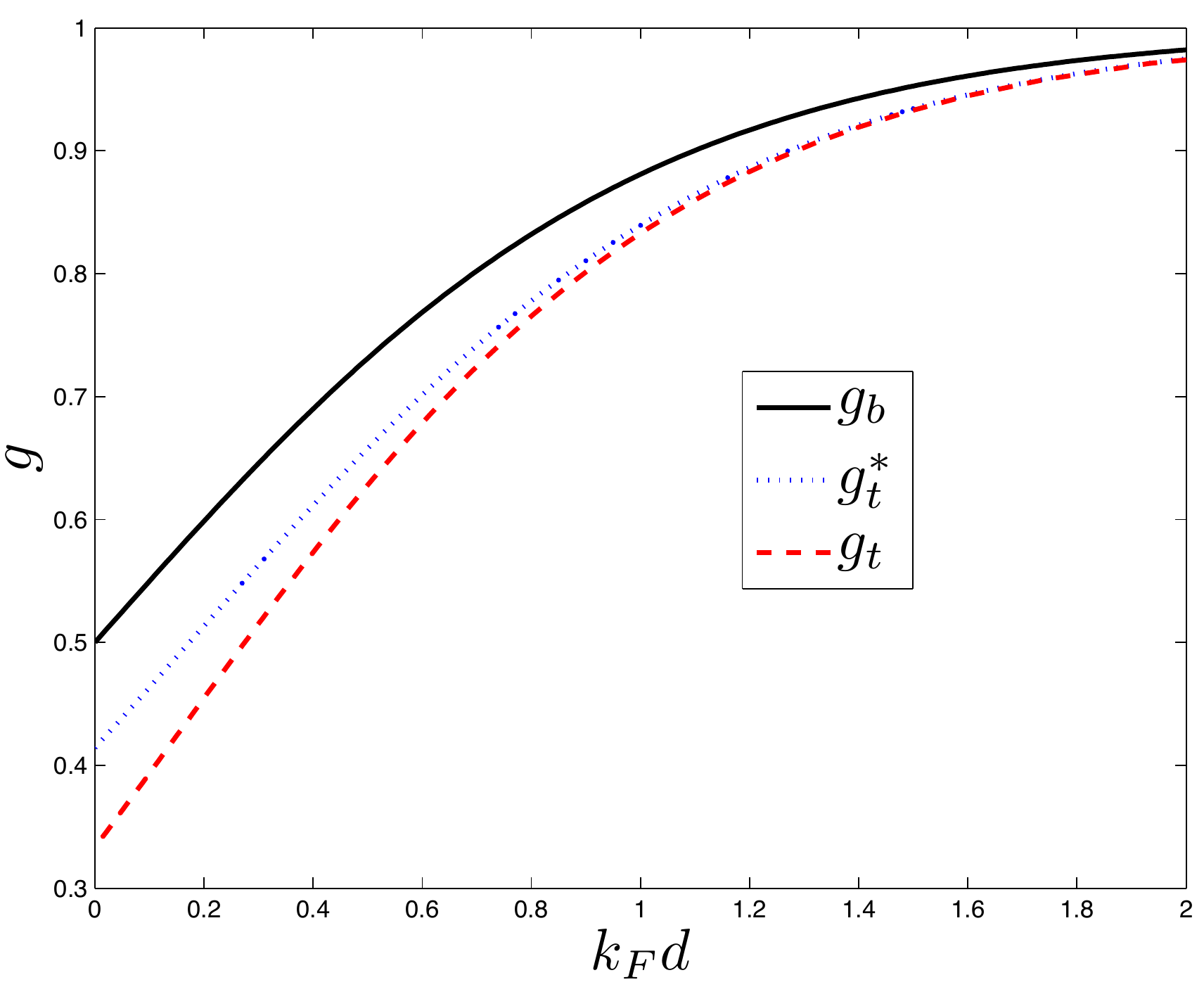,scale=0.45}
  \caption{Critical value for the appearance of a density-wave instability at $2k_F$ for $\theta=\cos^{-1}(\tfrac{1}{\sqrt{3}})$ and $\alpha=\pi/2$  excluding 
    exchange effects.
  The full (black) line is for a bilayer system, whereas the trilayer with all interaction is shown as a dashed (red) line and
  with only nearest-neighbor interaction as a dotted (blue) line.}
  \label{fig-crit}
\end{figure}

In Fig.~\ref{fig-crit} we plot the (lowest) bilayer and trilayer critical values at which the density-wave 
instability appears for $k_Fd\leq 2$. The single-layer critical value, $g_0=1$, is approached 
asymptotically, however, for the range plotted the multilayer cases are all below that 
value by at least a few percent. As expected the bilayer is always above the trilayer value. 
For $k_Fd\rightarrow 0$, the critical value undergoes
the largest reduction which is a factor of two for the bilayer, while for the trilayer it 
is a factor of 3 (we return to this fact below). We also compare 
the trilayer with and without the interaction of the two outer layers, $V_2(q)$. When excluding
the term, we see a larger critical value, $g^{*}_{t}$, for all $k_Fd$ than when taking it
into account in $g_t$. The additional attraction of $V_2(q)$ thus reduces the critical value as 
one would expect.

\subsection{Multiple Layers}
For more than three layers we expect similar behavior as seen above, i.e. a critical strength that decreases
with decreasing $k_Fd$. In the limit $k_Fd\rightarrow 0$ we can in fact find the exact solution for $g_N$ 
for any number of layers, $N$. Here the matrix in Eq.~\eqref{det} simplifies considerable since it 
has $1-g$ in all diagonal and $-g$ in all the non-diagonal entries. It is easy to verify that 
a vector consisting of ones in every entry  is an eigenvector of this matrix with eigenvalue $1-Ng$. 
We thus conclude that the system is unstable towards the formation of in-phase density waves in all planes 
for the critical coupling strength $g_N=1/N$. The limit of very small $k_Fd$ should therefore
approach this simple value. This limit is clearly seen for the lowest critical value 
in the bi- and trilayer cases above with the corresponding eigenmodes approaching one in all entries.
If we take this limit by reducing $d$ while keeping $k_F$ constant, we see that 
for large $N$ the critical value approaches zero as the layers come closer. Here the (non-interacting) 
system is equivalent to that of $N$ spins moving in two dimensions. The response function Eq.~\eqref{response} 
is then multiplied by a factor of $N$ which reduces the critical value by a factor $1/N$.

\subsection{Exchange effects}\label{exchange}
The exchange correction has to be included in the diagonal terms of the response function only as discussed in the
previous section. 
This means  that the eigenvectors corresponding to the critical couplings are 
the same {\it irrespective} of whether the exchange effect is included or not. 
In the large $k_Fd$ limit, the 
off-diagonal terms of the interaction are negligible, and we thus obtain the critical value $g=1/(1-G(2k_F))$.
Without exchange the single-layer result is recovered, i.e. $g=1$. Using the value $G(2k_F)=1/2$ as estimated
in the previous section, the limit is a factor of two larger. These arguments make it clear that the 
effects of exchange are more pronounced in limit of large $k_Fd$ where the intralayer correlations dominate. 
However, the interlayer correlations dominate for small $k_Fd$ which means that the exchange effects are insignificant 
in this limit. This means that one of our main results, the $1/N$ scaling of the critical coupling strength for $k_Fd\ll 1$, still holds 
when exchange is included.  

\begin{figure}[htb!]
  \epsfig{file=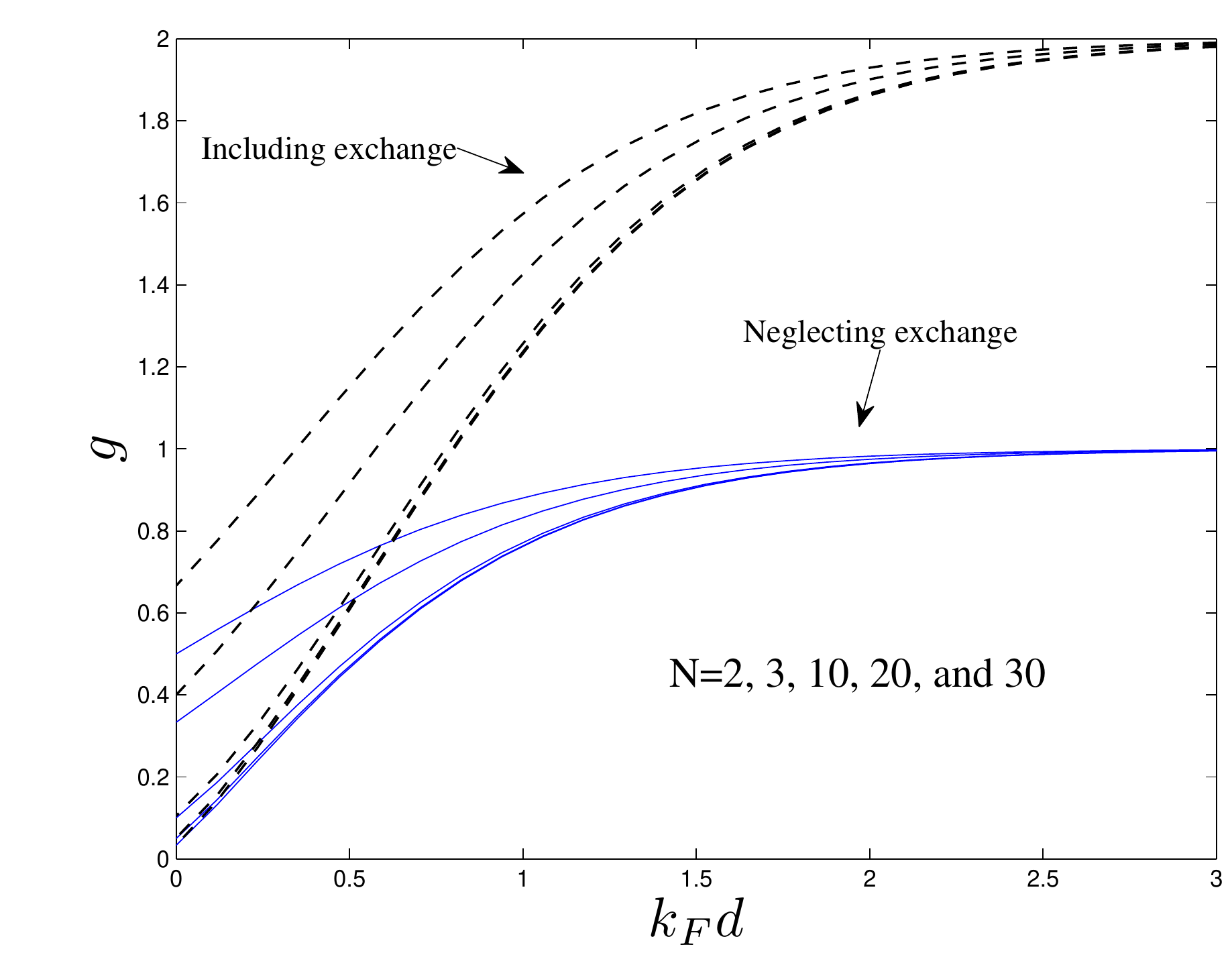,scale=0.45}
  \caption{Same as Fig.~\ref{fig-crit} but for $N=2$, 3, 10, 20, and 30 layers from top to bottom. Notice the limit
  at $k_Fd\rightarrow 0$ which is $g_N\rightarrow 1/N$.}
  \label{multi}
\end{figure}

In Fig.~\ref{multi} we show numerical solutions for the lowest critical values for $N=2$, 3, 10, 20, and 30, when
neglecting (lower full (blue) lines) and including exchange (upper dashed (black) lines). The expected 
decrease of $g_N$ with $N$ is clearly seen both with and without exchange corrections. 
For example, at $k_Fd=1$ the critical value is 0.88 for
$N=2$, whereas for $N=30$ it is 0.77 when neglecting exchange. The numbers are 1.57 for $N=2$ and 1.23 for $N=30$
when including exchange. This trend continues for higher $N$. Note that all the limits discussed above are clearly
confirmed by the numerics. In particular the results with and without including the local field factor to account for 
exchange approach each other as $k_Fd$ becomes small and when $N$ grows.

The corresponding eigenmodes 
all have the density waves in the layers in-phase. There are also other solutions with larger critical coupling strengths 
as for the bi- and trilayer cases. The eigenmodes can be analyzed in similar fashion and 
one finds that the solutions can be organized according to the number of adjacent layers that are out of phase 
with each other with the lowest solution (plotted in Fig.~\ref{multi}) fully in-phase across all layers.
We speculate that these higher modes correspond to collective modes in the striped phase. This will be  
examined in the future.

\subsection{Effective mass and Fermi surface deformation}\label{Deformation}
Finally, we briefly address the question of influence of effective mass and Fermi surface deformation caused by the 
dipolar interaction. For the single-layer case, these corrections have been calculated in Ref.~\cite{yamaguchi2010}; it was found that   
these terms pushes the critical value up by about 25\% for $\cos^2 \theta=1/3$
and   about  18\% for  $\theta=0$. In our setup this factor 
has to be included as a prefactor of $\chi^0$ in Eq.~\eqref{RPA}, i.e. it effectively amounts
a redefinition of our $g$. The neglect of effective mass and deformation effects means
that our results represent lower bounds. Note again that the eigenvectors for the unstable modes are unaltered
by these corrections.

\section{Phase Diagram}
\begin{figure}[htb!]
  \epsfig{file=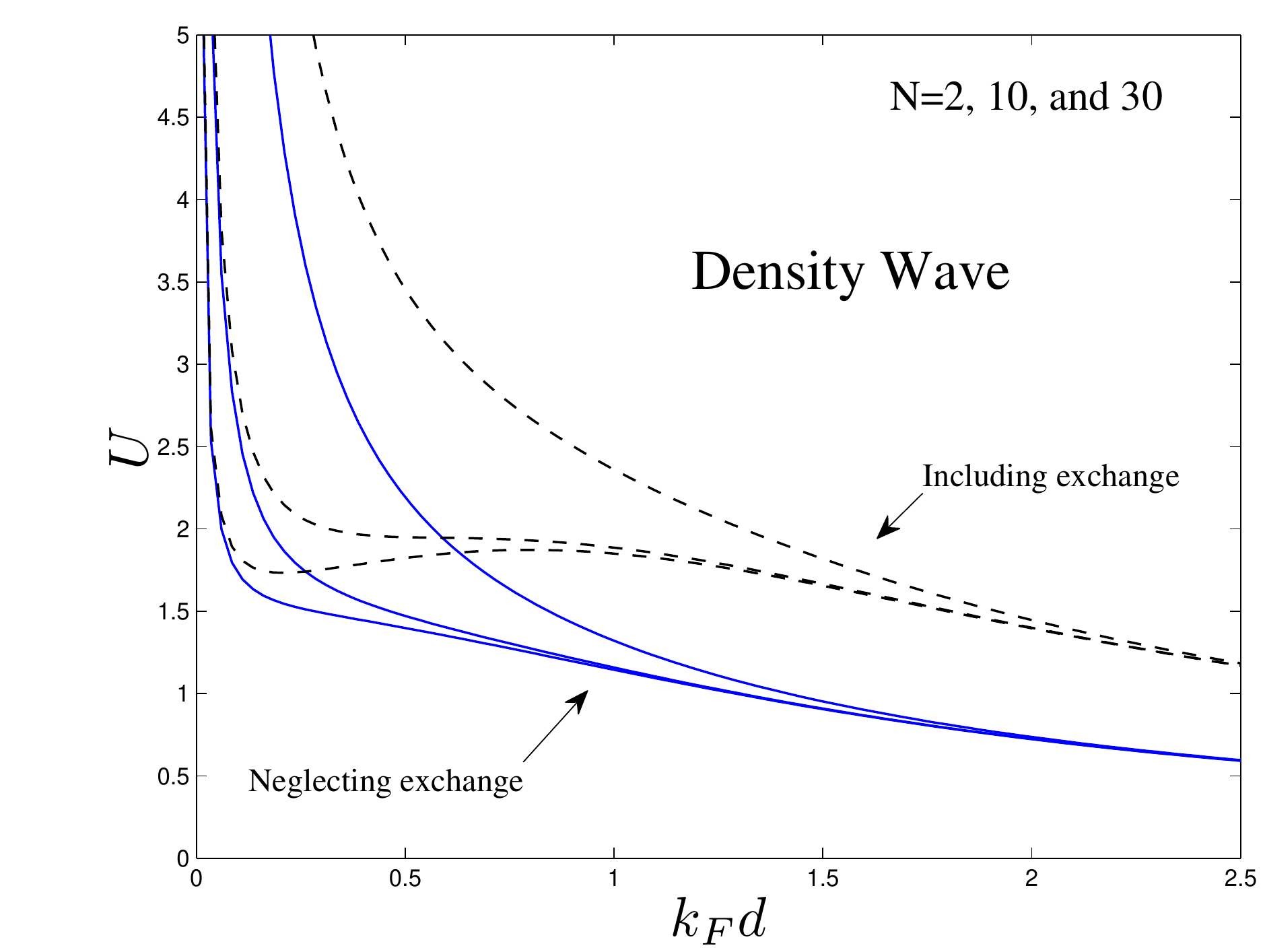,scale=0.45}
  \caption{Phase diagram at $\theta=\cos^{-1}(\tfrac{1}{\sqrt{3}})$ and $\alpha=\pi/2$ as function of $U=mD^2/\hbar^2d$ and $k_Fd$ for different number of layers $N=2$, 10, and 30. The density-wave instability occurs above the critical lines of which the full (blue) ones neglect while the dashed (black) ones take exchange effects into account.}
  \label{phase}
\end{figure}

In the multilayer setup, the interaction parameter, $U=mD^2/\hbar^2d$, is a convenient dimensionless measure
for the strength of interactions in the system. In Fig.~\eqref{phase} we show the zero temperature 
phase diagram in the $(U,k_Fd)$ plane for $N=2$, 10, and 30. The more layers, the earlier one expects to 
enter the density-wave regime as before. We also see that one can probe the phase diagram by changing
either the dipole moment or the density of fermions. Changing $d$ is also an option. This is, however, 
somewhat harder as $U\propto 1/d$ and the lines of constant $Uk_Fd=3g/2$ are very similar to the lines 
shown in Fig.~\eqref{phase}. We note that the inclusion of the exchange term causes an interesting 
plateau of the critical values for large $N$ at $k_Fd\sim 1$. This implies that there can be a large
region with $U\lesssim 2$ and $k_Fd\lesssim 1.5$ where the system is not unstable towards the formation
of density waves. This is valuable for the study of other phases like superfluidity which persists to 
small $U$ \cite{bruun2008}.

The regime of validity of the RPA approach augmented by the local field factor when applied to 
dipolar systems can be related to the corresponding situation for the electron liquid. 
In the latter case the RPA is known to provide reasonable results in the high density limit 
while it performs poorly at low densities where the Coulomb to kinetic energy ratio, $r_s$, becomes large \cite{giuli2005}. 
However, for dipolar systems the interaction dominates in the high density limit
whereas the low density limit is weakly interacting. We thus expect the RPA to be accurate for low densities and weak dipolar strengths,
i.e. when $g\ll 1$. This is precisely the case for the large $N$ limit which is our main interest in 
this work.

\subsection{Competing Phases and Finite Temperature}
The zero-temperature phase diagram for density-wave instabilities presented above needs to be 
considered in the light of other possible ground-states of the layered dipolar system. In the 
case of a single layer and in the weak couping limit, a $p$-wave superfluid state was proposed
\cite{bruun2008}. Likewise, a region of negative compressibility leading to 
collapse of the system was found \cite{bruun2008,yamaguchi2010}, although this happens outside
the parameter regime considered here. For several layers, the system can become superfluid with the 
Coopers pairs formed between dipoles residing in different layers~\cite{potter2010,pikovski2010,zinner2010,baranov2011}.
In the strong-coupling limit, a single layer of dipoles can also 
form a Wigner crystal with a symmetry which depends on how the dipoles are aligned with respect to the plane
~\cite{mora2007,buchler2007,astra2007,cremon2010}.
The presence of bound states in single and multilayer configuration of both fermionic 
and bosonic dipoles has also been given a great deal of attention recently 
\cite{shih2009,arm2010,klawunn2010,artem2010,fedorov2011,artem2011,wunsch2011}. For strong coupling, chains of
multiple dipoles in bound states could be the relevant degree of freedom in the system and 
the effective interaction of such constituents should determine the ground-state, and 
could be very different for odd fermionic chains as opposed to even bosonic ones. How
the phase diagram of single- and multilayer system at zero temperature maps out is 
an extremely interesting topic for future research.

At finite temperature one expects the physics to be governed by the 
Berezinskii-Kosterlitz-Thouless transition (BKT) \cite{bkt}. In the bilayer 
case the BKT physics is contained in the pairing order parameter in the 
weak-coupling limit or in a condensate of bosonic 
dimers in the strong-coupling limit \cite{zinner2010}. For multiple
layers similar dimerized phases are expected that are governed by the
BKT transition \cite{potter2010}.
The universal 
relation for the critical temperature scales with the superfluid density 
As the latter is proportional to the total density for strong coupling, the low
density regimes can be difficult to access. We speculate that the interlayer
interactions could help stabilize the low-temperature phases of the system 
and in turn easier to access experimentally as compared to a single layer.
Again this is a topic for future research.

\section{Conclusions}
We have considered the density-wave instability of dipolar fermionic polar molecules confined to a stack of two-dimensional 
layers. As the number of layers increases we find a reduction of the critical strength to enter the 
density-wave regime at all densities. The corresponding density waves are in-phase in all the planes. 
In the low density limit the critical strength even approaches 
zero as the number of layers grow.

\begin{acknowledgement}
We thank M.~M. Parish for numerous discussion and for providing valuable references.
\end{acknowledgement}

\end{document}